# Magnonic Holographic Memory

F. Gertz[1], A. Kozhevnikov[2], Y. Filimonov[2], and A. Khitun[1]

[1] Electrical Engineering Department, University of California - Riverside, Riverside, CA, USA, 92521

[2] Kotel'nikov Institute of Radioengineering and Electronics of Russian Academy of Sciences, Saratov Branch, Saratov, Russia, 410019

Holography is a technique based on the wave nature of light which allows us to utilize wave interference between the object beam and the coherent background[1]. This interference pattern is called a "hologram", from the Greek words "holos"-the whole and "gramma" - the writing because it contains the whole information of the object. Holography is usually associated with images being made from light, however, this is only a narrow field of holography. The first holograms were designed for use with electron microscopes[2] and it wasn't until a decade later with the advent of the laser that optical holographic images were popularized[3]. The discovery and development of holography has brought to life a variety of novel technologies including holographic interferometry[4], interferometric microscopy[5], holographic security[6], and enriched different fields of science and art[7]. Multiple other fields have had major contributions made by using wave interference to produce holograms, including acoustic holograms[8] used in seismic applications, and microwave holography[9] used in radar systems. Holography has been also recognized as a future data storing technology with unprecedented data storage capacity and ability to write and read a large number of data in a highly parallel manner[10]. Here, we present a magnonic holographic memory device exploiting spin wave interference.

A spin wave is a collective excitation due to oscillation of electron spins in a magnetic lattice. Similar to phonons (lattice vibrational waves) spin wave propagation occurs in magnetic lattices where spins are coupled to each other via the exchange and dipole-dipole interactions and a quantized spin wave is referred to by a quasiparticle called a magnon. In contrast to optical waves, spin waves are characterized by relatively slow group velocity (e.g. about $10^5$-$10^8$ cm/s[11,12]) and short decay time (e.g. about $10^{-9}$-$10^{-6}$ s at room temperature[12]). These technological characteristics have left many to ignore spin waves as information carriers in the past. The situation has changed drastically as the characteristic size of modern logic devices has been scaled down to the deep nanometer range, where short propagation distances compensate the slow propagation speed and damping effects. This gave a new impetus to the development of spin wave-based devices, and several prototypes have been demonstrated[13-15]. The collective nature of a spin wave comprising a large number of coupled spins manifests itself in a long coherence length (e.g. tens of microns[16] or even millimeters[17] at room temperature), which makes it superior to other approaches in spintronics relying on a single spin [18-21]. That is the most appealing property of magnonics allowing us to rework the well-established optical techniques at the nanometer scale.

The concept of Magnonic Holographic Memory (MHM) for data storage and data processing has been recently proposed[22]. MHM evolves the general idea of optical approach to applications in the magnetic domain aimed to combine the advantages of magnetic data storage with the unique capabilities for read-in and read-out provided by spin waves. At the same time, the use of spin waves implies certain requirements on the memory design, which are mainly associated with the need to preserve the energy of the spin wave carrying signals and the mechanisms of spin wave excitation and detection. The schematics of MHM as described in Ref.[22] are shown in Figure 1(A). It comprises two major components: a magnetic matrix and an array of spin wave generating/detecting elements – input/output ports. Spin waves are excited by the elements on one or several sides of the matrix, propagate through the matrix and detected on the other side of the structure. For simplicity, the matrix is depicted as a two-dimensional grid of magnetic wires. These wires serve as a media for spin wave propagation – spin wave buses. The elementary mesh of the grid is a cross-junction between the two orthogonal magnetic wires as shown in Figure 1(A). There is a nano-magnet on the top of each junction. Each of these nano-magnets is a memory element holding information encoded in the magnetization state. The nano-magnet can be designed to have two or several thermally stable states for magnetization, where the number of states defines the number of logic bits stored in each junction. The spins of the nano-magnet are coupled to the spins of the junction magnetic wires via the exchange and/or dipole-dipole coupling affecting the phase of the propagation of spin waves. The phase change received by the spin wave depends on the strength and the direction of the magnetic field produced by the nano-magnet. At the same time, the spins of nano-magnet are affected by the local magnetization change caused by the propagating spin waves. We consider two modes of operations: read-in and read-out. In the read-in mode, the magnetic state of the junction can be switched if the amplitude of the transmitted spin wave exceeds some threshold value. In the read-out mode, the amplitudes of the propagating spin waves are too small to overcome the energy barrier between the states. So, the magnetization of the junction remains constant in the read-out mode.

MHM has multiple input/output ports located at the edges of the waveguides. These elements are aimed the convert the input electric signals into spin waves, and vice versa, convert the output spin waves into the electric signals. There are several possible ways of building such elements by using micro-antennas[16], spin torque oscillators[23], and multi-ferroic elements [24]. For example, a micro-antenna is the most widely used tool for spin wave excitation and detection in ferromagnetic films[25]. An electric current passed through the antenna placed in the vicinity of magnetic film generates a magnetic field around the current-carrying wires, which excites spin waves in the magnetic material. And vice versa, a propagating spin wave changes the magnetic flux from the magnetic waveguide and generates the inductive voltage in the antenna contour.

The incident spin wave front is produced by the number of spin wave generating elements (e.g. by the elements on the left side of the matrix as illustrated in Figure 1(B)). All the elements are biased by the same RF generator exciting spin waves of the same frequency $f$ and amplitude $A_0$, while the phase of the generated waves are controlled by the DC voltages applied individually to the each element. Thus, the elements constitute a phased array allowing us to artificially change the angle of illumination by providing a phase shift between the input waves.

Propagating though the junction, spin wave accumulates an additional phase shift $\Delta\phi$, which depends on the strength and the direction of the local magnetic field provided by the nano-magnet $H_m$:

$$\Delta\phi = \int_0^r k(\vec{H}_m) dr \quad ,$$

where the particular form of the wavenumber $k(H)$ dependence vary for magnetic materials, film dimensions, and the mutual direction of wave propagation and the external magnetic field[26]. For example, spin waves propagating perpendicular to the external magnetic field (magnetostatic surface spin wave – MSSW) and spin waves propagating parallel to the direction of the external field (backward volume magnetostatic spin wave – BVMSW) may obtain significantly different phase shifts for the same field strength. The phase shift $\Delta\phi$ produced by the external magnetic field variation $\delta H$ in the ferromagnetic film can be expressed as follows[13]:

$$\frac{\Delta\phi}{\partial H} = \frac{l}{d}\frac{(\gamma H)^2 + \omega^2}{2\pi\gamma^2 M_s H^2} \text{ (BVMSW)}, \quad \frac{\Delta\phi}{\delta H} = -\frac{l}{d}\frac{\gamma^2(H + 2\pi M_s)}{\omega^2 - \gamma^2 H(H + 4\pi M_s)} \text{ (MSSW)},$$

where $\Delta\phi$ is the phase shift produced by the change of the external magnetic field $\delta H$, $l$ is the propagation length, $d$ is the thickness of the ferromagnetic film, $\gamma$ is gyromagnetic ratio, $\omega = 2\pi f$, $4\pi M_s$ is saturation magnetization of the ferromagnetic film.

The output voltage is a result of superposition of all the excited spin waves traveled though the different paths through the matrix. The amplitude of the output voltage is within the ultimate value corresponding to the maximum when all the waves are coming in-phase (constructive interference), and the minimum when the waves cancel each other (destructive interference).

In this work, we present the first realization of a 2-bit MHM as shown in Figure 2. The magnetic matrix is a double-cross structure made of yttrium iron garnet $Y_3Fe_2(FeO_4)_3$ (YIG) epitaxially grown on gadolinium gallium garnet $Gd_3Ga_5O_{12}$ substrate with (111) crystallographic orientation. YIG film has ferromagnetic resonance (FMR) linewidth $2\Delta H \approx 0.5 Oe$, saturation magnetization $4\pi M_s = 1750G$, and thickness $d=3.6\mu m$. This material is chosen due to its long spin wave coherence length and relatively low damping[27], which makes it the best candidate for room temperature spin wave devices prototyping. The length of the whole structure is 3mm, the width of the arm in 360µm. There are two micro-magnets on the top of the cross junctions. These magnets are the memory elements, where logic bits are encoded into the two possible directions for magnetization. There are six micro-antennas fabricated on the top of the YIG waveguides. These antennas are used to excite spin wave in YIG material and to detect the inductive voltage produced by the propagating spin waves.

A schematic of the experiment is shown in Figure 2(A). The input and the output micro-antennas are connected to the Hewlett-Packard 8720A Vector Network Analyzer (VNA). The VNA generates an input RF signal up to 20 GHz and measures the S parameters showing the amplitude and the phases of the transmitted and reflected signals. The prototype is placed inside an electro-magnet allowing variation in the bias magnetic field from -1000Oe to +1000Oe. The input from VNA is split between the four inputs via the two splitters, where the amplitudes of the signals are equalized by the attenuators (step $\pm 1dB$). The phases of the signal provided to the ports 3 and 4 are controlled by the two phase shifters ($\pm 2^0$). The photo of the YIG structure is shown in Figure 2(B).

The graph in Figure 2(C) shows the raw data collected for the structure with just two working micro-antennas. The test experiments are provided for the structure without magnet placed on the junctions. The graph shows the amplitude of the output inductive voltage detected for different excitation frequencies in the range from 5.30GHz to 5.55GHz. The curves of different color depict the output obtained for different phase difference $\Delta\phi$ among the two inputs 2 and 3. These data show the oscillation of the output voltage as a function of frequency and the phase difference between the two generated spin waves. The frequency dependence of the output is attributed to the effect of spin wave confinement within the structure, while the phase-dependent oscillations reveal the interference nature of the output signal. In Figure 2 (D), we show the slice of the data taken at the fixed frequency of 5.42GHz. The experimental data has a good fit with the classical equation for the two interfering waves. The only notable discrepancy is observed for $\Delta\phi=\pi$, where experimental value is non-zero. This fact can be well understood by taking into account all possible parasitic effects (e.g. reflecting waves, direct coupling between the input/output ports, structure imperfections, etc.)

Next, we carried out the experiments placing two micro-magnets with a length of 1.1mm and a width of 360µm, each with a proprietary micro-scale coating provided by Paramount Sensors and a coercivity of 200-500 Oe on the junctions as illustrated in Figure 3. The aim of these experiments is to show the dependence of the output from the magnetic states of the micro-magnets. Figure 3 shows the set of three holograms obtained for the three configurations of the top micro-magnets as illustrated by the schematics: A) two micro-magnets aligned in the same direction perpendicular to the long axis; B) the magnets are directed in the orthogonal directions; and C) both magnets are directed along the long axis. The red markers show the experimentally measured data (inductive voltage in millivolts) obtained at different phases of the four generated spin waves. The cyan surface is a computer reconstructed 3-D plot.  The excitation frequency is 5.40 GHz, the bias magnetic field is 1000 Oe. All experiments are done at room temperature. As one can see from Figure 3, the state of the micro-magnet significantly changes the output. The three holograms clearly demonstrate the unique signature defined by the magnetic state of the micro-magnet. The internal state of the holographic memory can be reconstructed by the difference in amplitude as well as the phase-dependent distribution of the output.

There are several important observations we want to outline based on the obtained experimental data.  The observed interference data are matched well by the classical equation (as shown in Figure 2(D)). The only feasible deviation is at the point corresponding to the destructive interference, where the experimental data show a non-zero output due to the presence of the additional parasitic coupling (e.g. waveguide eigen modes, reflecting waves, direct coupling among the input/output antennas, etc). Such a good fit shows no feasible sign of the thermal noise. The latter can be explained by taking into account that the flicker noise level in ferrite structures usually does not exceed -130 dBm[28]. At the same time, the direct coupling between the input/output antennas in our case is of the order of -70-80 dBm. So, all possible thermal noise signals are much smaller than the direct coupling, reflections and others parasitic effects.

Phase-dependent output is clearly observed in a relatively long device, where the direct distance from the input to the output exceeds 3mm. In contrast to the initial skepticism on the possibility of using spin wave interference for logic applications, it appears to be a robust instrument allowing us to sense magnetic textures at micrometer scale at room temperature.   Long coherent is the main physical parameter defining the maximum size of the interference-based devices. Taking into account the typical size of currently used nano-magnetic bits (i.e. ~100nm x 100 nm [29]), spin waves make it possible to sense hundreds of thousands of bits in parallel.

The main challenge with further magnonic hologram scaling down is associated with the development of nano-elements for spin wave excitation and detection. The use of micro-antennas implies certain limits on the minimum detector size as the output inductive voltage is proportional to the area of the detecting conducting contour. Spin torque oscillators[23] or multiferroic elements[30] are among the most promising candidates for I/O elements. It was recently demonstrated a two-terminal spin wave device where the excitation and detection of the spin wave signal was accomplished by the pair of multiferroic elements comprising magnetostrictive and piezoelectric materials [24]. The main advantages of using two-phase multiferroics are ultra-low power consumption and negligible coupling between the input/output elements, which eliminates the problems inherent to the micro-antenna approach.

In this work, we have demonstrated the first magnonic holographic memory device utilizing spin waves for information read-out from the magnetic matrix with two micro-magnets. Our results show the feasibility of applying the holographic techniques in magnetic structures, combining the advantages of magnetic data storage with the wave-based information transfer. Though spin waves cannot compete with photons in terms of the propagation speed and exhibit much higher losses, magnonic holographic devices may be more suitable for nanometer scale integration with electronic circuits. Another potential advantage of the spin wave approach is that the operating wavelength can be scaled down to the nanometer scale, which translates in the possibility of increasing data storage density to 1Tb/cm$^2$. The development of magnonic holographic memory devices and their incorporation within integrated circuits may pave a road to the next generation of logic devices exploiting phase in addition to amplitude for logic functionality.

### Acknowledgments

This work was supported in part by the FAME Center, one of six centers of STARnet, a Semiconductor Research Corporation program sponsored by MARCO and DARPA and by the National Science Foundation under the NEB2020 Grant ECCS-1124714.

**Figure Captions**

Figure 1.(A) The schematics of Magnonic Holographic Memory consisting of a 4×4 magnetic matrix and an array of spin wave generating/detecting elements For simplicity, the matrix is depicted as a two-dimensional grid of magnetic wires with just 4 elements on each side. These wires serve as a media for spin wave propagation. The nano-magnet on the top of the junction is a memory element, where information is encoded into the magnetization state. The spins of the nano-magnet are coupled to the spins of the magnetic wires via the exchange coupling. (B) Illustration of the principle of operation. Spin waves are excited by the elements on one or several sides of the matrix (e.g. left side), propagate through the matrix and detected on the other side (e.g. right side) of the structure. All input waves are of the same amplitude and frequency. The initial phases of the input waves are controlled by the generating elements. The output waves are the results of the spin wave interference within the matrix. The amplitude of the output wave depends on the initial and the magnetic states of the junctions.

Figure 2. (A) The schematics of the experimental setup. The test under study is a double-cross YIG structure with six micro-antennas fabricated on the edges. The input and the output micro-antennas are connected to the Hewlett-Packard 8720A Vector Network Analyzer (VNA). The VNA generates input RF signal in the range from 5.3GHz to 5.6GHz and measures the S parameters showing the amplitude and the phases of the transmitted and reflected signals. (B) The photo of the YIG double-cross structure. The length of the structure is 3mm, and the arm width is 360μm. (C) Transmitted signal $S_{12}$ spectra for the structure without micro-magnets. Two input signals are generated by the micro-antennas 2 and 3. The curves of different color show the output inductive voltage obtained for different phase difference among the two interfering spin waves. (D) The slice of the data taken at the fixed frequency of 5.42GHz (black curve). The red curve shows the theoretical values obtained by the classical equation for the two interfering waves.

Figure 3. A set of three holograms obtained for the three configurations of the top micro-magnets as illustrated by the schematics on the top: A) two micro-magnets aligned in the same direction perpendicular to the long axis; B) the magnets are directed in the orthogonal directions; and C) both magnets are directed along the long axis. The red markers show the experimentally measured data (inductive voltage in millivolts) obtained at different phases of the four generated spin waves. The cyan surface is a computer reconstructed 3-D plot. The excitation frequency is 5.4GHz, the bias magnetic field is 1000 Oe, all experiments are done at room temperature.

# References

1  Gabor, D. Nobel Lecture, Holography, 1948-1971. *Nobel Lectures, Physics 1971-1980, Editor Stig Lundqvist, World Scientific Publishing Co., Singapore* (1992).
2  Gabor, D. Microscopy by reconstructed wavefronts. *Proceedings of the Royal Society of London, Series A (Mathematical and Physical Sciences)* **197**, 454-487 (1949).
3  HANDBOOK OF OPTICAL HOLOGRAPHY - CAULFIELD,HJ. *Leonardo* **22**, 444-444 (1989).
4  Jones, R. & Wykes, C.   (ed Cambridge University Press) (1989).
5  Schwarz, C. J., Kuznetsova, Y. & Brueck, S. R. J. Imaging interferometric microscopy. *Optics Letters* **28**, 1424-1426 (2003).
6  Su, W.-C., Chen, Y.-W., Chen, Y.-J., Lin, S.-H & Wang, L.-K. Security optical data storage in FOurier holograms. *Applied Optics* **51**, 1297-1303 (2012).
7  Thompson, B. J. *Applications of holography*.  (1971).
8  Brenden & Hildebrande. *An Introduction to Acoustical Holography*.  (Plenum Press, 1972).
9  Kock, W. E. *Engineering Applications of Lasers and Holography*.  (Plenum Press, 1975).
10  *Data Storage*.  (2010).
11  Fletcher, P. C. & Kittel, C. Considerations on the Propogation and Generation of Magnetostatic Waves and Spin Waves. *Physical Review* **120**, 2004-2006 (1960).
12  Olson, F. A. & Yaeger, J. R. Propogation, Dispersion, and Attenuation of Backward-Traveling Magnetostatic Waves in YIG. *Applied Physics Letters* **5**, 33-35 (1964).
13  Kostylev, M. P., Serga, A. A., Schneider, T., Leven, B. & Hillebrands, B. Spin-wave logical gates. *Applied Physics Letters* **87**, 153501-153501-153503 (2005).
14  Wu, Y. *et al.* A Three-Terminal Spin-Wave Device for Logic Applications. *Journal of Nanoelectronics and Optoelectronics* **4**, 394-397, doi:10.1166/jno.2009.1045 (2009).
15  Schneider, T. *et al.* Realization of spin-wave logic gates. *Appl. Phys. Lett.* **92**, 022505-022503 (2008).
16  Covington, M., Crawford, T. M. & Parker, G. J. Time-resolved measurement of propagating spin waves in ferromagnetic thin films. *Physical Review Letters* **89**, 237202-237201-237204 (2002).
17  Serga, A. A., Chumak, A. V. & Hillebrands, B. YIG magnonics. *Journal of Physics D-Applied Physics* **43**, doi:264002 10.1088/0022-3727/43/26/264002 (2010).
18  Wolf, S. A., Awschalom, D. D., Buhram, R. A. & Daughton, J. M. Spintronics: A spin-based electronics vision for the future. *Science* (2001).
19  Flatte, M. E. & Awschalom, D. D. Challenges for semiconductor spintronics. *Nature Physics* (2007).
20  Parkin, S. S. P. & Bader, S. D. Spintronics. *Annual Review of Condensed Matter Physics* (2010).
21  Zutic, I., Fabian, J. & Sarma, S. D. Spintronics: Fundamentals and Applications. *Reviews of Modern Physics* (2004).
22  Khitun, A. Magnonic holographic devices for special type data processing. *JOURNAL OF APPLIED PHYSICS* **113**, doi:164503 10.1063/1.4802656 (2013).
23  Kaka, S. *et al.* Mutual phase-locking of microwave spin torque nano-oscillators. *Nature* **437**, 389-392 (2005).
24  Cherepov, S. *et al.* Electric-field-induced spin wave generation using multiferroic magnetoelectric cells. *Proceedings of the 56th Conference on Magnetism and Magnetic Materials (MMM 2011), DB-03, Scottsdale, Arizona* (2011).
25  Silva, T. J., Lee, C. S., Crawford, T. M. & Rogers, C. T. Inductive measurement of ultrafast magnetization dynamics in thin-film Permalloy. *Journal of Applied Physics* **85**, 7849-7862 (1999).
26  Eschbach, J. & Damon, R. *J. Phys. Chem. Solids* **19**, 308 (1961).

27	Serga, A. A., Chumak, A. V. & Hillebrands, B. YIG MAgnonics. *Journal of Physics D: Applied Physics* **43** (2010).
28	Rubiola, E., Gruson, Y. & Giordano, V. On the flicker noise of ferrite circulators for ultra-stable oscillators. *Ieee Transactions on Ultrasonics Ferroelectrics and Frequency Control* **51**, 957-963, doi:10.1109/tuffc.2004.1324399 (2004).
29	Imre, A. *et al.* Majority Logic Gate for Magnetic Quantum-Dot Cellular Automata. *Science* **311**, 205-208 (2006).
30	Shabadi, P. *et al.* Towards logic functions as the device. *2010 IEEE/ACM International Symposium on Nanoscale Architectures (NANOARCH 2010)*, doi:10.1109/nanoarch.2010.5510934 (2010).

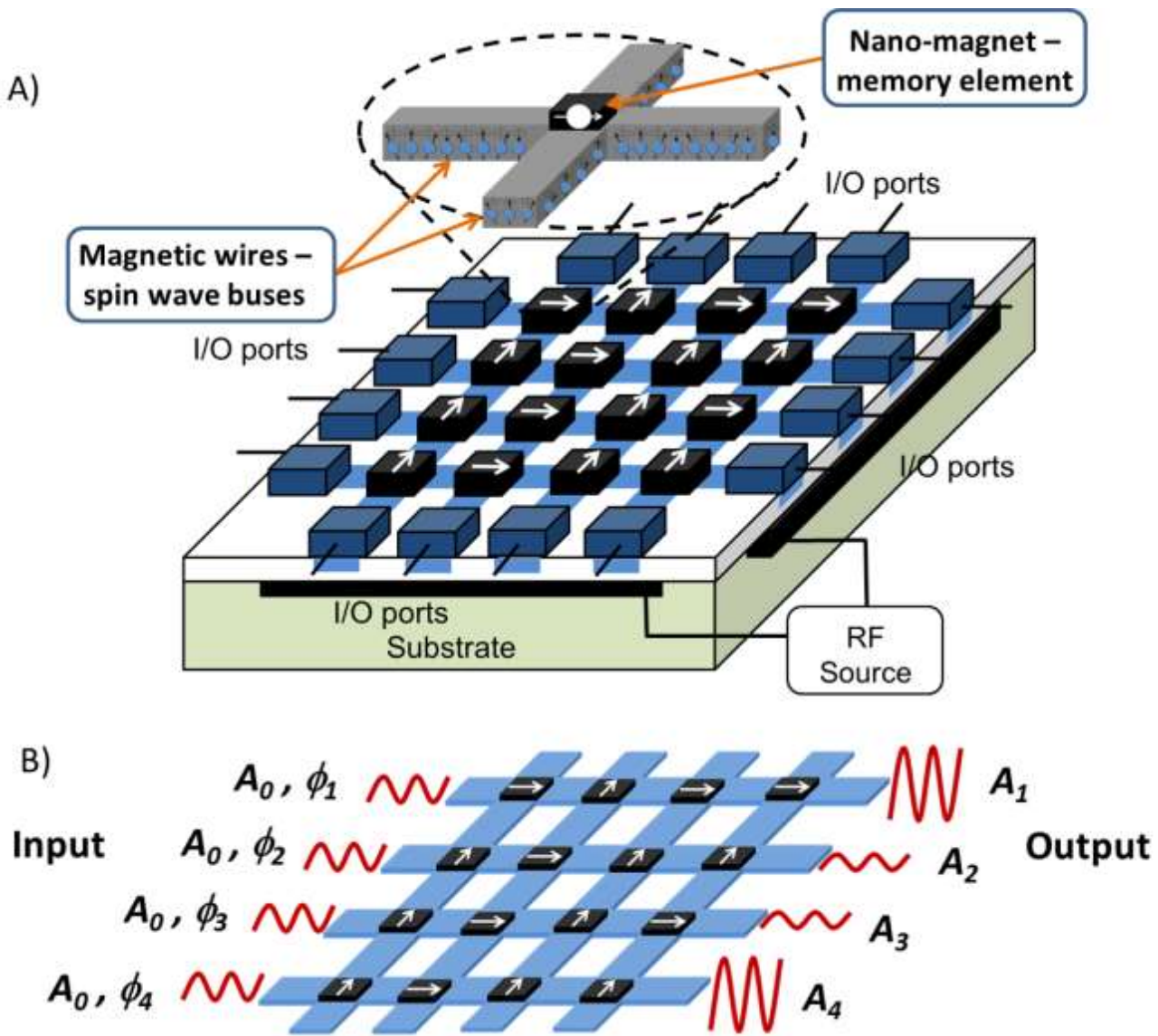

Figure 1

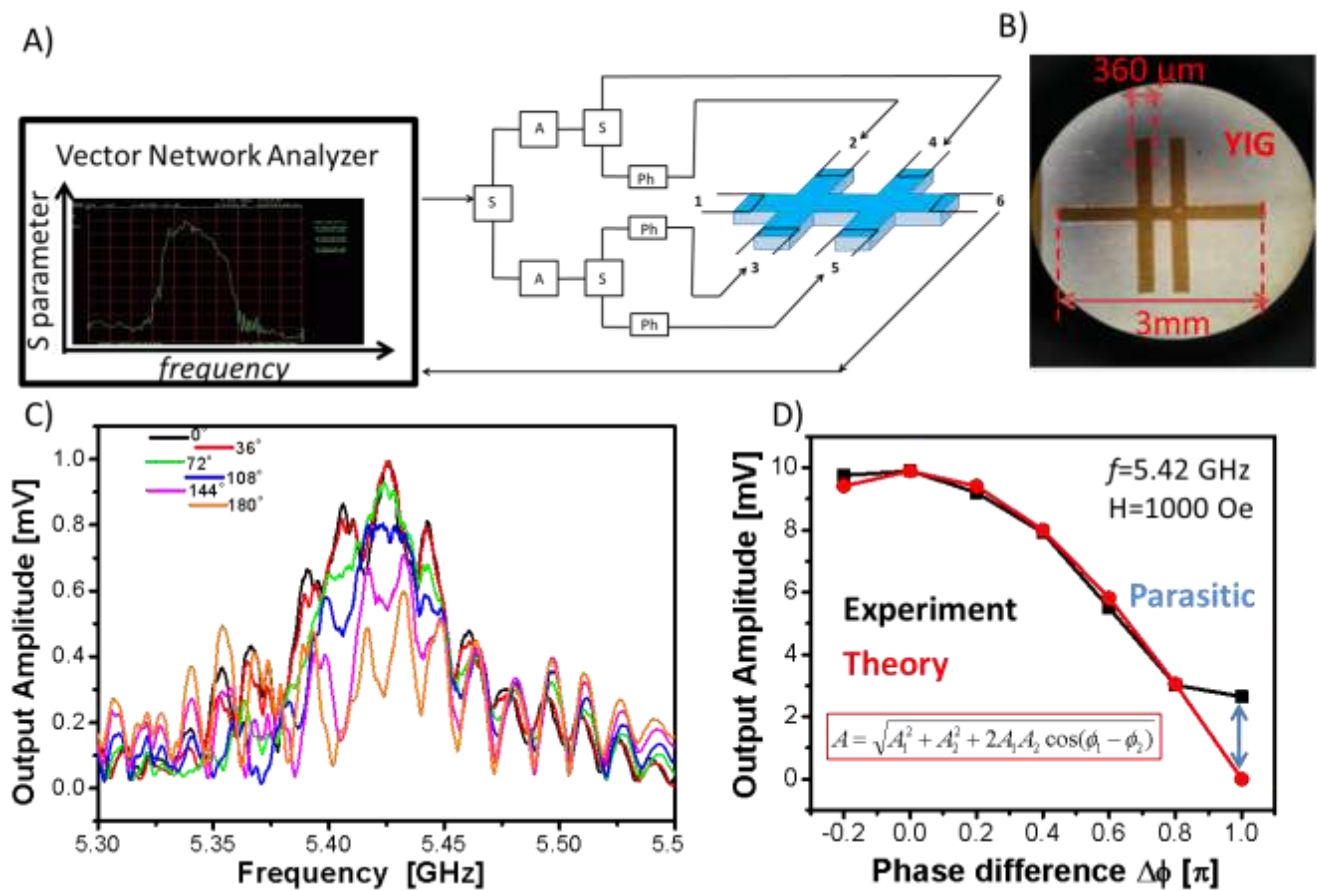

Figure 2

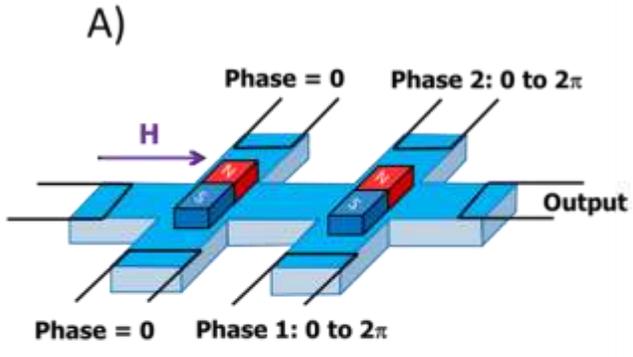
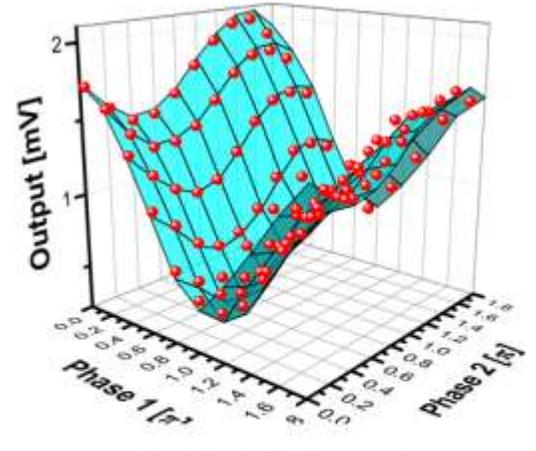
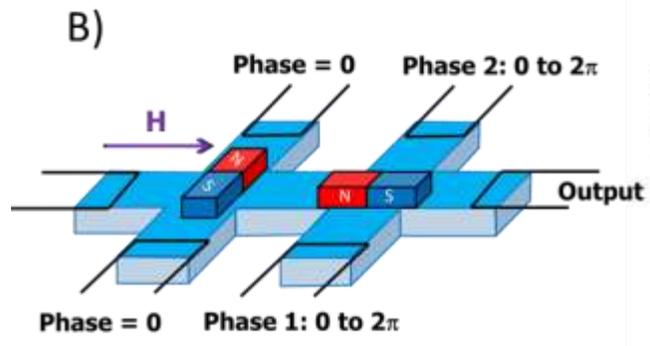
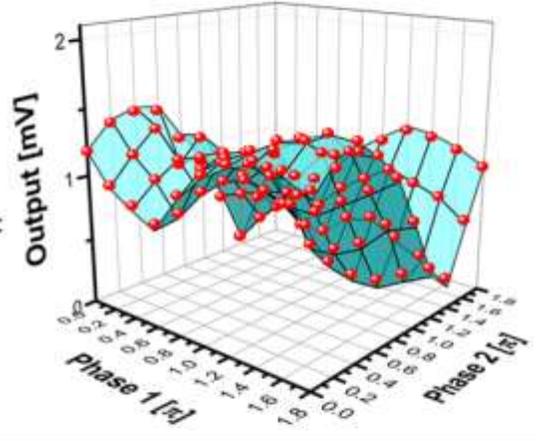
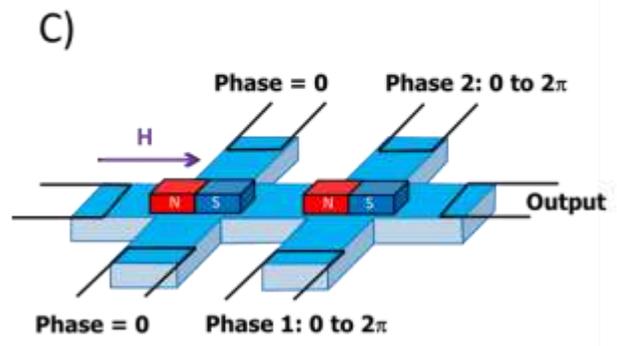
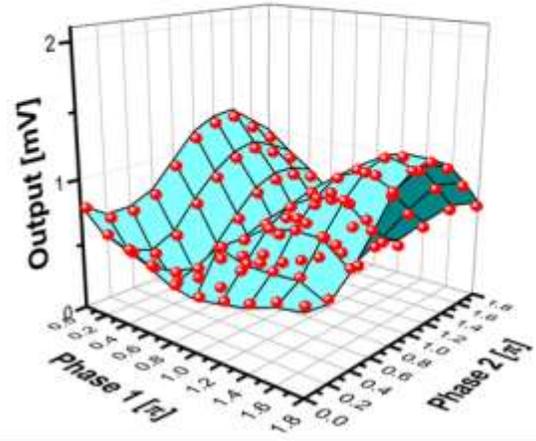

Figure 3